\documentclass{ws-procs9x6}
\begin{document}

\title{RICCI FLOW DEFORMATION\\
OF COSMOLOGICAL INITIAL DATA SETS}

\author{M. CARFORA$^\diamond$ and T. BUCHERT$^\dagger $}

\address{$^\diamond $Dipartimento di Fisica Nucleare e Teorica, Universita' degli Studi di Pavia,\\
and\\
Istituto Nazionale di Fisica Nucleare, Sezione di Pavia,\\
Pavia, Via Bassi 6, 27100, Italy\\
E-mail: mauro.carfora@pv.infn.it}
\address{$^\dagger $Universit$\acute e$ Lyon 1, Centre de Recherche Astrophysique de Lyon, CNRS UMR 5574,  \\ 9 avenue Charles Andr$\acute e$, F--69230 Saint--Genis--Laval, France\\
E-mail: buchert@obs.univ-lyon1.fr }

\begin{abstract}
Ricci flow deformation of cosmological initial data sets in general
relativity is a technique for generating families of initial data sets
which potentially would allow to interpolate between distinct spacetimes.
This idea has been around since the appearance of the Ricci flow on the
scene, but it has been difficult to turn it into a sound mathematical
procedure. In this expository talk we illustrate, how Perelman's recent
results in Ricci flow theory can considerably improve on such a situation. 
From a physical point of view this analysis can be related to the issue of
finding a constant--curvature template spacetime for the inhomogeneous Universe,
relevant to the interpretation of observational data and, hence, 
bears relevance to the dark energy and dark matter debates. These techniques 
provide control on curvature fluctuations (intrinsic backreaction terms)
in their relation to the averaged matter distribution.
\end{abstract}


\bodymatter

\section{INTRODUCTION}

The Ricci flow has been introduced by R. Hamilton \cite{11} with the goal of
providing an analytic approach to  Thurston's geometrization conjecture for
three-manifolds \cite{thurston1,thurston2}$\,$. Inspired by the theory of harmonic maps,
he considered the geometric evolution equation obtained when one evolves a
Riemannian metric $g_{ab}$, on a three-manifold $\Sigma $, in the direction of its
Ricci tensor\cite{Huisken} $\mathcal{R}_{ab}$, \emph{i.e.}
\begin{equation}
\left\{ 
\begin{tabular}{l}
$\frac{\partial }{\partial \beta }g_{ab}(\beta )=-2\,\,\mathcal{R}_{ab}(\beta )\;,$ \\ 
\\ 
$g_{ab}(\beta =0)=g_{ab}$\, ,\;\; $0\leq \beta <T_{0}\;$.%
\end{tabular}
\right.   \label{mflow}
\end{equation}
\noindent In recent years, this geometric flow has gained  extreme popularity thanks
to the revolutionary breakthroughs of G. Perelman \cite{18, 19, 20}$\,$, who, taking the
whole subject by storm,  has brought to completion Hamilton's approach to Thurston's
conjecture. The prominent themes recurring in Hamilton's and Perelman's works
converge to a proof that  the Ricci flow, coupled to topological surgery, provides a
natural technique for factorizing and uniformizing a three-dimensional Riemannian
manifold $(\Sigma,g)$  into locally homogeneous geometries. This is a result of vast
potential use also in theoretical physics, where the Ricci flow often appears in
disguise as a natural real-space renormalization group flow. Non-linear $\sigma
$-model theory, describing quantum strings propagating in a background spacetime,
affords the standard case study  in such a setting \cite{9, {car1}, lottCMC, Bakas,
Bakas2, Oliynyk}$\,$. 

Another paradigmatical, perhaps even more direct, application occurs in relativistic
cosmology \cite{CarfMarz2, CarfKam}$\,$, (for a series of recent results see also
\cite{4,5} and the references cited therein). This will be related to the main topic
of this talk, and to motivate our interest in it, let us recall that 
homogeneous and isotropic solutions of Einstein's laws of gravity (the
Friedman--Lemaitre--Robertson--Walker (FLRW) spacetimes) do not account for
inhomogeneities in the
Universe. The question whether they do {\it on average} is an issue \cite{ellis}
that is the subject of considerable debate especially in the recent literature (see
\cite{kolbetal, rasanen} 
and follow--up references; comprehensive
lists may be found in \cite{ellisbuchert, rasanenrev} and \cite{Buchrev}). 

In any case, a member of the family of FLRW cosmologies (the so--called concordance
model that is characterized by a dominating cosmological
constant in a spatially flat universe model) provides a successful {\em  fitting model}
to a large number of observational data, and the generally held view is that the
spatial sections of FLRW spacetimes indeed describe the {\em physical
Universe} on a sufficiently large averaging scale. This raises an interesting
problem in mathematical cosmology: devise a way to explicitly construct a
constant--curvature metric out of a \emph{scale--dependent}  inhomogeneous
distribution of matter and spatial curvature. It is in such a framework that one
makes the basic observation that the Ricci flow (\ref{mflow}) and its linearization,
provide a natural technique\cite{CarfKam, 4}  for deforming, and under suitable
conditions smoothing,  the geometrical part of scale--dependent cosmological initial
data sets. Moreover, by taking advantage of some elementary aspects of Perelman's
results,
this technique also provides a natural and unique way for deforming, along the Ricci
flow, the matter distribution. The expectation is that in this way we can define a
deformation of cosmological initial data sets into a one-parameter family of initial
data whose time evolution, along the evolutive part of Einstein's equations, 
describe the Ricci flow deformation of a cosmological spacetime.

\section{The kinematical set--up: Initial data set for cosmological spacetimes}

\medskip\noindent
To set notation, we emphasize that throughout the paper we shall consider a smooth
three-dimensional manifold $\Sigma $, which we assume to be closed and without
boundary. We let $C^{\infty }(\Sigma, \mathbb{R})$ and 
${C}^{\infty }(\Sigma,\otimes ^{p}\, T^{*}\Sigma\otimes ^{q}T\Sigma )$ be the space
of smooth functions and of smooth $(p,q)$--tensor fields on $\Sigma $, respectively.
 
We shall denote by $\mathcal{D}iff(\Sigma )$ the group of smooth diffeomorphisms of
$\Sigma $, and by $\mathcal{R}iem(\Sigma )$ the space of all smooth Riemannian
metrics over $\Sigma$. The tangent space\,, $\mathcal{T}_{(\Sigma
,g)}\mathcal{R}iem(\Sigma )$, to $\mathcal{R}iem(\Sigma )$ at $(\Sigma,g )$ can be
naturally identified with the space of symmetric bilinear forms ${C}^{\infty
}(\Sigma,\otimes ^{2}\, T^{*}\Sigma)$ over $\Sigma $.  The hypothesis of smoothness
has been made for simplicity. 
Results similar to those described below,  can be obtained for
initial data sets with finite Holder or Sobolev differentiability.
\noindent In such a framework, let us recall that a collection  of  fields $g\in
\mathcal{R}iem(\Sigma )$, 
$K\in \mathcal{T}_{(\Sigma ,g)}\mathcal{R}iem(\Sigma )$, $\varrho \in C^{\infty
}(\Sigma, \mathbb{R}^{+})$, $\vec{J}\in {C}^{\infty }(\Sigma, T\Sigma )$, defined
over the three-manifold $\Sigma $, characterizes 
\ a set  $(\;\Sigma \;,\;g_{ab}\;,\;K_{ab}\;,\;\varrho \;,\;J_{a}\;)$, 
of physical cosmological initial data for Einstein equations if and only if the
\emph{matter fields} $(\varrho ,\vec{J})$ verify the weak energy condition $\varrho
\geq 0$, the dominant energy condition $\varrho ^{2}\geq g_{ab}J^{a}J^{b}$,  and
their coupling with the geometric fields $(g,K)$ is such as to satisfy the
Hamiltonian and divergence constraints\footnote{%
Latin indices run through $1,2,3$; we adopt the summation convention. The
nabla operator denotes covariant derivative with respect to the 3--metric.
The units are such that $c=1$.}: 
\begin{eqnarray}
\mathcal{R}+{k}^{2}-K_{\;\,b}^{a}K_{\;\,a}^{b} &=&16\pi G\varrho +2\Lambda
\;,  \label{constraint1} \\
\nabla _{b}K_{\;\,a}^{b}-\nabla _{a}k &=&8\pi GJ_{a}\;.  \label{constraints}
\end{eqnarray}
Here $\Lambda $ is the cosmological constant, $k:=g^{ab}K_{ab}$, and  $%
\mathcal{R}$ \ is the scalar curvature of the Riemannian metric $g_{ab}$. If such a
set of admissible data is
propagated according to the evolutive part of Einstein's equations, then the
symmetric tensor field $K_{ab}$ can be interpreted as the extrinsic
curvature and $k$ as the mean curvature of the embedding $i_{t}:\Sigma
\rightarrow M^{(4)}$ of $(\Sigma ,g_{ab})$ in the spacetime $(M^{(4)}\simeq
\Sigma \times \mathbb{R},g^{(4)})$ resulting from the evolution of $(\Sigma
,g_{ab},K_{ab},\varrho ,J_{a})$, whereas $\varrho $ and $J_{a}$ are,
respectively, identified with the mass density and the momentum density of
the material self--gravitating sources on $(\Sigma ,g_{ab})$. 

\section{The Heuristics of averaging: Deformation of cosmological initial data sets}

\medskip\noindent
The averaging procedure described in \cite{CarfMarz2,CarfKam,4} is based on a
 smooth deformation of the physical initial data  $(\Sigma
,g_{ab},K_{ab},\varrho ,J_{a})$ into a one-parameter family of initial data
sets 
\begin{equation}
\beta \longmapsto (\Sigma \;,\;g_{ab}(\beta )\;,\;K_{ab}(\beta )\;,\;\varrho
(\beta )\;,\;J_{a}(\beta )),  \label{betadata}
\end{equation}
with $0\leq \beta \leq \infty $ being a parameter characterizing the averaging
scale. The general idea is to construct the flow (%
\ref{betadata}) in such a way as to represent, as $\beta $ increases, a
scale-dependent averaging of $(\Sigma ,g_{ab},K_{ab},\varrho ,J_{a})$,
and -- under suitable hypotheses -- reducing it to a constant-curvature initial data
set 
\begin{equation}
(\Sigma \;,\;\overline{g}_{ab}\;,\;\overline{K}_{ab}=\frac{1}{3}\overline{g}%
_{ab}\overline{k}\;,\;\overline{\varrho }\;,\;\overline{J}_{a}=0),
\end{equation}
where $\overline{g}_{ab}$ is a constant curvature metric on $\Sigma $, $%
\overline{k}$ is the (spatially constant) trace of the extrinsic curvature
(related to the Hubble parameter), and $\overline{\varrho }$ is the averaged
matter density. Under the heading of such a general strategy it is easy to figure
out the reasons for an important role played by the Ricci flow  and its
linearization.  As we shall recall shortly, they are  natural geometrical flows
always defining a non-trivial deformation of the metric $g_{ab}$ and of the
extrinsic curvature $K_{ab}$. Moreover, when global, they posses remarkable
smoothing properties. For instance, if, for $\beta =0$, the scalar curvature
$\mathcal{R}$ of $(\Sigma ,g_{ab})$
is $>0$, and if there exist positive constants $\alpha _{1}$, $\alpha _{2}$, 
$\alpha _{3}$, not depending on $\beta $, such that 
$\mathcal{R}_{ab}(\beta )-\alpha _{1}g_{ab}(\beta )\mathcal{R}(\beta )\geq 0$,\
and  $\widehat{\mathcal{R}}_{ab}(\beta )\widehat{\mathcal{R}}^{ab}(\beta )\leq
\alpha _{2}\mathcal{R}^{1-\alpha _{3}}(\beta )$,\
where $\widehat{\mathcal{R}}_{ab}(\beta )\doteq \mathcal{R}_{ab}(\beta )-\frac{1}{3}%
g_{ab}(\beta )\mathcal{R}(\beta )$\
is the trace--free part of the Ricci tensor, then \cite{11} the solutions 
$ (g_{ab}(\beta ), K_{ab}(\beta ))$ of the (volume--normalized) Ricci flow and its
linearization  exist for all $\beta >0$, and the pair $(%
g_{ab}(\beta ),K_{ab}(\beta ))$,  uniformly converges, when $\beta \rightarrow
\infty $, to $(\overline{g}_{ab},\mathcal{L}_{\vec{v}}\overline{g}_{ab})$ where
$\overline{g}_{ab}$ is a metric with constant positive sectional
curvature,  and $\vec{v}$ is some vector field on $\Sigma $, possibly depending on
$\beta$. The flow $\beta \mapsto
(\overline{g}_{ab},\mathcal{L}_{\vec{v}}\overline{g}_{ab})$, describing a motion by
diffeomorphisms over a constant curvature manifold, can be thought of as
representing the smoothing of the geometrical part of an initial data set $(\Sigma
,g_{ab},K_{ab},\varrho ,J_{a})$.  \\

\noindent It is useful to keep in mind what we can expect and what we cannot expect
out of such a Ricci--flow deformation of cosmological initial data set. Let us start
by remarking that the family of data (\ref{betadata}) will correspond to the initial
data for physical spacetimes iff the constraints 
(\ref{constraint1}) and (\ref{constraints}) hold throughout the $\beta $--dependent
deformation. This is a very strong requirement and, if we have a technically
consistent way $\beta \mapsto(\varrho (\beta ), g_{ab}(\beta ), K_{ab}(\beta ))$ of
deforming the matter distribution and the geometrical data, then  the most natural
way of implementing the constraints is to use them  to define scale--dependent
backreaction fields $\beta \mapsto \phi (\beta )$, $\beta \mapsto \psi _{a} (\beta
)$  describing the non--linear interaction between matter averaging and geometrical
averaging, \emph{i.e.},

\begin{eqnarray}
\phi(\beta ) &\doteq &  \varrho (\beta ) -(16\pi \,G)^{-1}\left[\mathcal{R}(\beta
)+{k}^{2}(\beta )-K_{\;\,b}^{a}(\beta )K_{\;\,a}^{b}(\beta )- 2\Lambda
\right]\;,\label{hamRo}\\
\psi _{a} (\beta ) &\doteq & J_{a}(\beta )-(8\pi \,G)^{-1} \left[\nabla
_{b}K_{\;\,a}^{b}(\beta )-\nabla _{a}k(\beta )   \right]\;.
\label{divJ}
\end{eqnarray} 

\noindent 
To illustrate how this strategy works, let us concentrate, in this talk, on the
characterization of the scalar field $\beta \mapsto \phi (\beta )$, providing the
backreaction between matter and geometrical averaging. The covector field $\beta
\mapsto \psi _{a} (\beta )$ can, in principle, be controlled by the action of a
$\beta $--dependent diffeomorphism. However, its analysis requires a subtle
interplay with the kinematics of spacetime foliations \cite{buchert:grgfluid}$\,$,
(\emph{i.e.}, how we deal with the lapse function and with the shift vector field in
the framework of Perelman's approach), and will be discussed elsewhere, (for a
pre--Perelman approach to this issue see \cite{CarfKam, 4})$\,$. \\

\noindent Let us start by observing that the matter averaging flow $\beta \mapsto
\varrho (\beta )$  must comply with the preservation of the physical matter content 
\begin{equation}
\int_{\Sigma }\varrho (\beta )\, d\mu _{g(\beta )}=   \int_{\Sigma }\varrho (\beta=0
)\, d\mu _{g(\beta=0 )}\doteq M\,,\;\;\forall \beta \;, 
\label{locMass}
\end{equation}
and must be explicitly coupled to the scale of geometrical averaging. In other
words, if, for some fixed $\beta>0 $, we consider that part of the matter
distribution $\varrho (\beta )$ which is localized in a given region $B(x,\tau
(\beta) )\subset \Sigma _{\beta }$ of size $\tau (\beta)$, then we should be able to
tell from which localized distribution $(\varrho _{m},B(x,\tau ))$, at $\beta =0$, 
the selected  matter content   $(\varrho (\beta ),B(x,\tau (\beta)))$ has evolved. 
A natural answer to these requirements is provided by Perelman's backward
localization\cite{18} of probability measures on Ricci evolving manifolds.    The
idea  is to probe the Ricci flow with a probability measure whose dynamics can
localize the regions of the manifold $(\Sigma, g)$ of geometric interest.
This is achieved by  considering, along the solution  $g_{ab}(\beta )$ of
(\ref{mflow}),  a $\beta $--dependent mapping 
$\beta \longmapsto f(\beta,\;)\in C^{\infty }(\Sigma_{\beta} ,\mathbb{R})$, in
terms of which one constructs on $\Sigma_{\beta}$ the  measure $d\varpi (\beta
)\doteq \left( 4\pi \tau (\beta )\right) ^{-\frac{3}{2}}e^{-f}d\mu _{g(\beta )}$,
where $\beta \longmapsto \tau (\beta )\in \mathbb{R}^{+}$
is a scale parameter chosen in such a way as to normalize $d\varpi (\beta )$
according to the so--called \emph{Perelman's coupling}\,:\
$\int_{\Sigma_{\beta} }d\varpi (\beta )=\left( 4\pi \tau (\beta )\right) ^{-\frac{3}{2}}\int_{\Sigma_{\beta} }e^{-f}d\mu _{g(\beta )}=1$. \
It is easily verified that this is preserved in form along the Ricci flow
(\ref{mflow}),
if the mapping $f$ and the scale parameter $\tau (\beta )$ are evolved backward in
time $\beta \in (\beta ^{*},0)$ according to the coupled flows defined by 
\begin{equation}
\left\{ 
\begin{tabular}{l}
$\frac{\partial }{\partial \beta }f=-\Delta _{g(\beta)}f-R(\beta )+\frac{3}{2}\tau
(\beta )^{-1},$  $f(\beta ^{*})=f_{0}$ \\ 
\\ 
$\frac{d }{d \beta }\;\tau (\beta )=-1,$ $\tau(\beta^{*})=\tau_{0},$%
\end{tabular}
\right.   \label{ourp}
\end{equation}
where $\Delta _{g(\beta )}$ is the Laplacian with respect to the
metric $g_{ab}(\beta )$, and $f_{0}$, $\tau_{0}$ are given (final) data.
In this connection, note that
the equation for $f$ is a
backward heat equation, and as such the forward evolution $f(\beta
=0)\rightarrow f$ is an ill-posed problem. A direct way for circumventing such a
difficulty is to 
interpret (\ref{ourp}) according to the
following two-steps prescription: (i) Evolve the metric $\beta \longmapsto (\Sigma
,g_{ab}(\beta ))$, say up
to some $\beta ^{*}$, according to the  Ricci flow, (if the flow is global we may
let $\beta
^{*}\rightarrow \infty $);
(ii) On the Ricci evolved Riemannian manifold $(\Sigma ,\overline{g}%
_{ab}(\beta ^{*}))$ so obtained, select a function $f(\beta ^{*})$
and  the corresponding  scale parameter $\tau (\beta ^{*})$,\ \
 and evolve them,
backward in $\beta $, according to (\ref{ourp}). \\

\section{Ricci--flow deformation of the initial data $(\Sigma ,g_{ab}, K_{ab},
\varrho )$}

\medskip\noindent
With these preliminary remarks along the way, let us characterize the various steps
involved in constructing the flow (\ref{betadata}). (For ease of exposition, we refer to the standard unnormalized flow; volume normalization can be enforced by a reparametrization of the deformation parameter). \\
\begin{definition} (\emph{Geometrical Data Deformation})
Given an initial data set $(\Sigma,g_{ab},K_{ab},\varrho,J_{a})$ for a cosmological
spacetime $(M^{(4)}\simeq \Sigma \times \mathbb{R},g^{(4)})$, the Ricci--flow
deformation of its geometrical part $(\Sigma,g_{ab},K_{ab})$ is defined by  the flow
$\beta \mapsto (g_{ab}(\beta ), K_{hj}(\beta))$, $0\leq  \beta <\beta ^{\ast}$
provided by the (weakly) parabolic initial value problem (the Ricci flow, proper) 
\begin{equation}
\left\{ 
\begin{tabular}{l}
$\frac{\partial }{\partial \beta }g_{ab}(\beta )=-2\mathcal{R}_{ab}(\beta ),$ \\ 
\\ 
$g_{ab}(\beta =0)=g_{ab}$\, ,\;\; $0\leq \beta <\, \beta ^{\ast}\;$\\
\\
$ \lim_{\beta \nearrow \beta ^{\ast} }\left[\sup_{x\in \Sigma }|Rm(x,\beta )|
\right]<\infty    \;,$%
\end{tabular}
\right.   \label{Fflow}
\end{equation}
and by its linearization, which, by suitably fixing the action of the diffeomorphism
group $\mathcal{D}iff(\Sigma )$,  takes the form of the parabolic initial value
problem \cite{chowluni}
\begin{equation} 
\left\{ 
\begin{tabular}{l}
$\frac{\partial  }{\partial  \beta   }{K}_{ab}(\beta  )=\Delta _{L}\, {K}_{ab}(\beta)\;,$ \\ 
\\ 
${K}_{ab}(\beta  =0)=\,K_{ab}$\, ,\;\; $0\leq \beta  <\beta^{\ast }$\;,%
\end{tabular}
 \right.  \label{linDT0}
\end{equation}
where $\Delta _{L}$ denotes the Lichnerowicz--DeRham Laplacian $\Delta
_{L}{K}_{ab}\doteq \nabla ^{i}\nabla _{i} {K}%
_{ab}-R_{as}{K}_{b}^{s}-R_{bs}{K}_{a}^{s}+2R_{asbt}{K%
}^{st}$ acting on symmetric bilinear forms \cite{lichnerowicz}$\,$. 
\end{definition}
\begin{definition} (\emph{Localization of the Deformed Data})
The geometrical deformation $\beta \mapsto (g_{ab}(\beta ), K_{hj}(\beta))$ is
controlled by the backward localizing flow
$\eta \mapsto (E(x,y;\eta), E_{i'k'}^{ab}(x,y;\eta))$,\ $\eta \doteq \beta ^{\ast
}-\beta $, defined by the backward heat kernels for the conjugate
\cite{Glickenstein, carflin} parabolic initial value problem associated with the
$g(\eta)$--dependent Laplace-Beltrami and Lichnerowicz operators $\eta \mapsto
(\Delta,\,\Delta_{L} )$. The operator $\Delta_{L} $, when acting on (bi)scalars,
reduces to $\Delta $. Thus, we can characterize both these kernels  in a compact
form as solutions of  
\begin{equation} 
\left\{ 
\begin{tabular}{l}
$\frac{\partial }{\partial \eta }\;\dbinom{{E_{i'k'}^{ab}(x,y;\eta)}}{{E(x,y;\eta)}}
=\Delta_{L}^{(x)}\;\dbinom{{E_{i'k'}^{ab}(x,y;\eta)}}{{E(x,y;\eta)}}-
\,\mathcal{R}\;\dbinom{{E_{i'k'}^{ab}(x,y;\eta)}}{{E(x,y;\eta)}}\;,$\\
\\
$\lim_{\;\eta \searrow
0^{+}}\;\dbinom{{E_{i'k'}^{ab}(x,y;\eta)}}{{E(x,y;\eta)}}=\dbinom{{\delta
_{i'k'}^{ab}(x,y;\eta)}}{{\delta (x,y;\eta)}}\;,$%
\end{tabular}
\right. \;   \label{fund}
\end{equation} 
where $(y,x;\eta )\in (\Sigma \times \Sigma \backslash Diag(\Sigma \times \Sigma
))\times [0,\beta ^{*}]$,   $\Delta_{L}^{(x)}$ denotes the  Lichnerowicz--DeRham
Laplacian with respect to the variable $x$, the heat kernels
${E}^{ab}_{i'k'}(y,x;\eta )$\, and $E(x,y;\eta)$  are smooth sections of $(\otimes
^{2}T\Sigma)\boxtimes (\otimes ^{2}T^{*}\Sigma)$ and $\Sigma\boxtimes \Sigma$,
respectively, and finally, ${\delta }^{k_{1}...k_{p}}_{i'_{1}...i'_{p}}(y,x;\eta)$
is the Dirac $p$--tensorial measure, ($p=0,1,\ldots$) on $(\Sigma, g(\eta ))$. 
The Dirac initial condition is understood in the distributional sense, \emph{i.e.}, 
$\int_{\Sigma _{\eta }}{K}^{ab}_{i'k'}(y,x;\eta )\;w^{i'k'}(y)\;d\mu^{(y)} _{g(\eta
)}\rightarrow w^{ab}(x)\;\;\;as\;\;\eta \searrow 0^{+}$, for any smooth symmetric
bilinear form with compact support $w^{i'k'}\in C^{\infty }_{0}(\Sigma ,\otimes
^{2}T\Sigma )$, and where the limit is meant in the uniform norm on  $C^{\infty
}_{0}(\Sigma ,\otimes ^{2}T\Sigma )$. 
\end{definition}
Note that
heat kernels  for generalized Laplacians, such as $\Delta  _{L}$, (smoothly)
depending on a one--parameter family of metrics $\varepsilon \mapsto
g_{ab}(\varepsilon )$, $\varepsilon \geq 0$, are briefly dealt with in
\cite{vergne}$\,$. The delicate setting where the parameter dependence is, as in our
case, identified with the parabolic time driving the diffusion of the kernel, is
discussed  in \cite{guenther2, chowluni}$\,$, (see Appendix A, \S 7 for a
characterization of the parametrix of the heat kernel in such a case), and in
\cite{lanconelli}$\,$. Strictly speaking, in all these works, the analysis is confined
to the scalar Laplacian, possibly with a potential term, but the theory readily
extends to generalized Laplacians, always under the assumption that the metric
$g_{ab}(\beta )$ is smooth as $\nearrow \beta ^{*}$. Finally, the kernels for
$\Delta$ and $\Delta_{L}$ can both be normalized, along the Ricci flow, over the
round (collapsing) 3--sphere $(\mathbb{S}^{3},g_{can}(\beta ))$.

\medskip

\noindent Let us now consider the matter content localized by taking the $d\varpi
(\eta  )$--expectation of $\varrho (\eta   )$, 
\begin{equation}
\int_{\Sigma }\,\varrho (\eta  )\,d\varpi (\eta )\doteq M(d\varpi (\eta ))\;.
\label{massaloc}
\end{equation}

\noindent According to (\ref{locMass}), we require that such a local mass is
preserved along the $\eta $--evolution of the measure $d\varpi (\eta )$,
\emph{i.e.}, $\frac{d}{d\beta }\int_{\Sigma }\,\varrho (\eta  )\,d\varpi (\eta
)=0$.
This request motivates the following
\begin{definition} (\emph{Deformation of Matter Data})
The given matter distribution $\varrho (\beta=0)$ is deformed according to the
heat--flow 
$\beta \mapsto \varrho (\beta)$ given by
\begin{equation}
\left\{ 
\begin{tabular}{l}
$\frac{\partial }{\partial \beta }\varrho (\beta )=\Delta _{g(\beta )}\,\, \varrho
(\beta ) , \,\, \beta\in [0,T_{0}) 
$ \\ 
\\ 
$\varrho (\beta =0)=\varrho.$%
\end{tabular}
\right.   \label{heatmass}
\end{equation}
\end{definition}

\noindent We are now in the position to characterize the Ricci flow deformation of
the cosmological data $(\Sigma,g_{ab},K_{ab},\varrho,J_{a})$. According to the
results described in \cite{carflin}  we have
\begin{proposition}
Let $\beta \mapsto (g_{ab}(\beta ),K_{ik}(\beta), \varrho (\beta ))$ be the Ricci
flow deformation, on $\Sigma_{\eta }\times [0,\beta ^{*}]$, of the data
$(\Sigma,g_{ab},K_{ab},\varrho)$ as defined above. Assume that the underlying Ricci
flow $\beta \mapsto (\Sigma ,g_{ab}(\beta ))$ is of bounded geometry, and  let
${E}(y,x;\eta )$ and  ${E}^{ab}_{i'k'}(y,x;\eta )$ be the (backward) heat kernels
Ricci-flow conjugated to the Laplace--Beltrami and to the Lichnerowicz--DeRham
operator, respectively. Then, for all $0\leq \eta \leq \beta ^{*}$,
\begin{equation}
 g_{i'k'}\,(y,\eta=0)=\int_{\Sigma
}{E}^{ab}_{i'k'}(y,x;\eta)\,\left[{g}_{ab}(x,\eta)-2\eta
\,\,\mathcal{R}_{ab}(x,\eta)\right]\,d\mu _{g(x,\eta )}\;,
\label{grepres0}
\end{equation}
\begin{equation}
\dbinom{{\mathcal{R}_{i'k'}(y,\eta=0 )}}{{{K}_{i'k'}(y,\eta=0 )}}
=\int_{\Sigma }{E}^{ab}_{i'k'}(y,x;\eta
)\,\dbinom{{\mathcal{R}_{ab}(x,\eta)}}{{{K}_{ab}(x,\eta)}}\,d\mu _{g(x,\eta )}
\;,
\label{riccikernel}
\end{equation}
\begin{equation}
and \qquad \varrho (y,\eta=0 )=\int_{\Sigma }{E}(y,x;\eta )\,\varrho (x,\eta)\,d\mu _{g(x,\eta
)}\;.
\end{equation}

\noindent Moreover, as $\eta\searrow 0^{+} $, we have the uniform asymptotic expansion

\begin{eqnarray}
&&\;\;\;\;\;\dbinom{{\mathcal{R}_{i'k'}(y,\eta=0 )}}{{{K}_{i'k'}(y,\eta=0 )}}\;=\;\nonumber\\
&&\frac{1}{\left(4\pi \,\eta \right)^{\frac{3}{2}}}\,\int_{\Sigma
}\exp\left(-\frac{d^{2}_{0}(y,x)}{4\eta } \right)\,{\tau }^{ab}_{i'k'}(y,x;\eta
)\,\dbinom{{\mathcal{R}_{ab}(x,\eta)}}{{{K}_{ab}(x,\eta)}}\,d\mu _{g(x,\eta
)}\nonumber\\
\nonumber\\
&&+\sum_{h=1}^{N}\frac{\eta ^{h}}{\left(4\pi \,\eta
\right)^{\frac{3}{2}}}\,\int_{\Sigma }\exp\left(-\frac{d^{2}_{0}(y,x)}{4\eta }
\right)\,{\Phi [h] }^{ab}_{i'k'}(y,x;\eta
)\,\dbinom{{\mathcal{R}_{ab}(x,\eta)}}{{{K}_{ab}(x,\eta)}}\,d\mu _{g(x,\eta
)}\nonumber\\
\nonumber\\
&&+O\left(\eta ^{N-\frac{1}{2}} \right)\;,
\end{eqnarray}
\noindent where ${\tau }^{ab}_{i'k'}(y,x;\eta )$ $\in T\Sigma_{\eta }\boxtimes
T^{*}\Sigma_{\eta } $  is the parallel transport operator
associated with  $(\Sigma ,g(\eta ))$,   $d_{0}(y,x)$ is the distance function in
$(\Sigma ,g(\eta=0 ))$, and ${\Phi [h] }^{ab}_{i'k'}(y,x;\eta )$ are smooth sections
 $\in C^{\infty }(\Sigma\times \Sigma ' ,\otimes ^{2}T\Sigma\boxtimes \otimes
^{2}T^{*}\Sigma)$, (depending on the geometry of $(\Sigma ,g(\eta ))$),
characterizing the asymptotics of the heat kernel ${E}^{ab}_{i'k'}(y,x;\eta )$. With
an obvious adaptation, such an asymptotic behavior also extends to
$g_{i'k'}\,(y,\eta=0)$ and $\varrho (y,\eta=0 )$.
\end{proposition}
With these results it is rather straightforward to provide useful characterizations
of the (intrinsic) backreaction field $\phi(\beta)$. For illustrative purposes, let
us assume that $\beta \mapsto K_{ab}(\beta )\equiv 0$ and $\Lambda \equiv 0$ , then
from (\ref{hamRo}) and the above proposition we get that
\begin{eqnarray}
&&\phi(y,\eta=0)=\,\int_{\Sigma }{E}(y,x;\eta )\,\varrho (x,\eta)\,d\mu _{g(x,\eta )}
\\
&&-(16\pi \,G)^{-1}\left[
\int_{\Sigma }g^{i'k'}(y,\eta=0)\,{E}^{ab}_{i'k'}(y,x;\eta
)\,{\mathcal{R}_{ab}(x,\eta)}\,d\mu _{g(x,\eta )}
 \right]\;.\nonumber
\end{eqnarray}
If we further assume that the Hamiltonian constraint holds at the \emph{fixed}
observative scale $\eta $ and that $\mathcal{R}_{ab}(x,\eta)\approx
[\frac{1}{3}\mathcal{R}(\eta )\,g_{ab}(x,\eta=0)+\delta \mathcal{R}_{ab}(x,\eta)]$,
(\emph{i.e.}, curvature fluctuates around the \emph{constant curvature background}
$\mathcal{R}_{l'm'}(y,\eta=0)=2C\,g_{l'm'}(y,\eta=0)$), then
\begin{equation}
\phi(y,\eta=0)\approx(16\pi \,G)^{-1}
\int_{\Sigma }g^{i'k'}(y,\eta=0)\,{E}^{ab}_{i'k'}(y,x;\eta )\,{\delta
\mathcal{R}_{ab}(x,\eta)}\,d\mu _{g(x,\eta )}
\;,
\end{equation}
from which it follows that the intrinsic backreaction field  $\phi(\beta)$ is
generated by curvature fluctuations around the given background, as expected.

\subsection*{Acknowledgements}
{\small M.C. would like to dedicate this paper to T. Ruggeri on occasion of his ..th birthday; he also thanks the organizers for a very stimulating and enjoyable meeting.
T.B. acknowledges hospitality at and support from the University of Pavia during a
working visit. Research supported in part by PRIN Grant $\# 2006017809$.}


\begin{thebibliography}{33}
\bibitem{Bakas2} I. Bakas, 
\textit{Renormalization group flows and continual Lie algebras}, 
J. High Energy Phys. 0308, 013 (2003).

\bibitem{Bakas} I. Bakas, 
\textit{Geometric flows and (some of) their physical applications},  
AvH conference Advances in Physics and Astrophysics of the 21st Century, 6-11
September 2005, Varna, Bulgaria, 
arXiv:hep-th/0511057 (2005).

\bibitem{vergne}
N. Berline, E. Getzler and M. Vergne, 
\textit{Heat kernels and Dirac operators},
Grundlehren Math. Wiss., vol. 298, Springer-Verlag, New York, (1992).

\bibitem{buchert:grgfluid}
T. Buchert, 
\textit{On average properties of inhomogeneous fluids in general relativity: 2. perfect fluid cosmologies},
Gen. Rel. Grav. {\bf 33}, 1381 (2001).

\bibitem{Buchrev} T. Buchert, 
\textit{Dark energy from structure: a status report},
Gen. Rel. Grav. (special issue on dark energy), in press; arXiv:0707.2153 (2007).

\bibitem{4}  T. Buchert and M. Carfora, 
\textit{Regional averaging and scaling in relativistic cosmology}, 
Class. Quant. Grav. \textbf{19}, 6109-6145 (2002).

\bibitem{5}  T. Buchert and M. Carfora, 
\textit{Cosmological parameters are dressed}, 
Phys.\ Rev.\ Lett. \textbf{90}, 31101-1-4 (2003). 

\bibitem{car1} M. Carfora, A. Marzuoli, 
\textit{Model geometries in the space of Riemannian structures and Hamilton's flow},
Class. Quantum Grav. 5, 659-693 (1988).  

\bibitem{CarfMarz2} M. Carfora and A. Marzuoli, 
\textit{Smoothing Out Spatially Closed Cosmologies}, 
Phys. Rev. Lett. \textbf{53}, 2445 (1984).

\bibitem{CarfKam} M. Carfora and K. Piotrkowska, 
\textit{Renormalization group approach to relativistic cosmology}, 
Phys. Rev. D \textbf{52}, 4393 (1995).

\bibitem{carflin} M. Carfora, 
\textit{The Conjugate Linearized Ricci Flow on Closed 3--Manifolds}, 
arXiv:0710.3342 (2007). 

\bibitem{Glickenstein} B. Chow, S-C. Chu, D. Glickenstein, C. Guenther, J. Isenberg,
T. Ivey, D. Knopf, P. Lu, L. Ni, 
\textit{The Ricci Flow: Techniques and Applications: Part I: Geometric Aspects},
Math. Surveys and Monographs Vol. \textbf{135}, Am. Math. Soc. (2007).

\bibitem{chowluni} B. Chow, P. Lu, L. Ni, 
\textit{Hamilton's Ricci Flow}, 
Graduate Studies in Math. Vol. \textbf{77}, Am. Math. Soc. (2007).

\bibitem{ellis} G.F.R. Ellis, 
\textit{Relativistic cosmology -- its nature, aims and problems},
in {\em General Relativity and Gravitation} (D. Reidel
Publishing Co., Dordrecht), pp.\ 215--288 (1984).

\bibitem{ellisbuchert} G.F.R. Ellis and T. Buchert, 
\textit{The Universe seen at different scales}, 
Phys. Lett. A (Einstein Special Issue)  \textbf{347}, 38 (2005).

\bibitem{9} D. H. Friedan, 
\textit{Nonlinear models in $2+\varepsilon$ dimensions}, 
Ann. Physics  \textbf{163}, no. 2, 318--419 (1985). 
 
\bibitem{lanconelli} N. Garofalo, E. Lanconelli, 
\textit{Asymptotic Behavior of Fundamental solutions and Potential Theory of Parabolic operators with variable coefficients}, 
Math. Ann. \textbf{283}, 211-239 (1989).
 
\bibitem{guenther2} C. Guenther, 
\textit{The fundamental solution on manifolds with time--dependent metrics}, 
J. Geom. Anal.  \textbf{12}, 425--436 (2002). 
 
\bibitem{11}  R.~S. Hamilton, 
\textit{Three--manifolds with positive Ricci curvature}, 
J. Diff. Geom. \textbf{17}, 255-306 (1982). 
 
\bibitem{Huisken} G. Huisken, 
\textit{Ricci deformation of the metric on a Riemannian manifold}, 
J. Differential Geom. \textbf{17}, 47--62 (1985).

\bibitem{kolbetal} E. W. Kolb, S. Matarrese and A. Riotto, 
\textit{On cosmic acceleration without dark energy}, 
New J. Phys., \textbf{8}, 322 (2006).

\bibitem{lichnerowicz}
A. Lichnerowicz, 
\textit{Propagateurs et commutateurs en relativit$\acute e$ g$\acute e$n$\acute e$rale.},   
Pub. Math. de  l'I.E.H.S. {\bf 10}, 5 (1961).

\bibitem{lottCMC} J. Lott, 
\textit{Renormalization group flow for general sigma models}, 
Comm. Math. Phys. {\bf 107}, 165--176 (1986). 
        
\bibitem{Oliynyk} T Oliynyk, V Suneeta, E Woolgar, 
\textit{A Gradient Flow for Worldsheet Nonlinear Sigma Models},
Nucl. Phys. B {\bf 739}, 441-458, arXiv:hep-th/0510239 (2006).         
        
\bibitem{18}  G. Perelman, 
\textit{The entropy formula for the Ricci flow and its geometric applications}, 
arXiv:math.DG/0211159 (2002).

\bibitem{19}  G. Perelman, 
\textit{Ricci flow with surgery on Three--Manifolds},
arXiv:math.DG/ 0303109 (2003).

\bibitem{20} G. Perelman, 
\textit{Finite extinction time for the solutions to the Ricci flow on certain three--manifolds}, 
arXiv:math.DG/0307245 (2003).

\bibitem{rasanen} S. R$\ddot a$s$\ddot a$nen, 
\textit{Dark energy from backreaction},
JCAP \textbf{042}, 003 (2004).

\bibitem{rasanenrev} S. R$\ddot a$s$\ddot a$nen 
\textit{Accelerated expansion from structure formation}, 
JCAP \textbf{0611}, 003 (2006).

\bibitem{thurston1} W. P. Thurston, 
\textit{Three-dimensional manifolds, Kleinian groups and hyperbolic geometry}, 
Bull. Amer. Math. Soc. (N.S.) \textbf{6}, 357-381 (1982).

\bibitem{thurston2} W. P. Thurston, 
\textit{Three-dimensiional geometry and topology}, 
Vol. 1. Edited by S. Levy. Princeton Math. Series, \textbf{35}, Princeton Univ. Press, Princeton NJ,  (1997).
\end{thebibliography}
\end{document}